# Segmentation of Knee Bones for Osteoarthritis Assessment: A Comparative Analysis of Supervised, Few-Shot, and Zero-Shot Learning Approaches

Yun Xin Teoh[1,2,§], Alice Othmani[2,§], Siew Li Goh[3,4,§], Juliana Usman[1], Khin Wee Lai[1]

[1]Department of Biomedical Engineering, Universiti Malaya, Kuala Lumpur, Malaysia
`teoh1648@gmail.com, juliana_78@um.edu.my, lai.khinwee@um.edu.my`
[2]LISSI, Université Paris-Est Créteil, Vitry sur Seine, France
`alice.othmani@u-pec.fr`
[3]Sports Medicine Unit, Faculty of Medicine, Universiti Malaya, Kuala Lumpur, Malaysia
`gsiewli@um.edu.my`
[4] Centre for Epidemiology and Evidence-Based Practice, Faculty of Medicine, Universiti Malaya, Kuala Lumpur, Malaysia

§ Corresponding author

**Abstract.** Knee osteoarthritis is a degenerative joint disease that induces chronic pain and disability. Bone morphological analysis is a promising tool to understand the mechanical aspect of this disorder. This study proposes a 2D bone morphological analysis using manually segmented bones to explore morphological features related to distinct pain conditions. Furthermore, six semantic segmentation algorithms are assessed for extracting femur and tibia bones from X-ray images. Our analysis reveals that the morphology of the femur undergoes significant changes in instances where pain worsens. Conversely, improvements in pain may not manifest pronounced alterations in bone shape. The few-shot-learning-based algorithm, UniverSeg, demonstrated superior segmentation results with Dice scores of 99.69% for femur and 99.60% for tibia. Regarding pain condition classification, the zero-shot-learning-based algorithm, CP-SAM, achieved the highest accuracy at 66% among all models. UniverSeg is recommended for automatic knee bone segmentation, while SAM models show potential with prompt encoder modifications for optimized outcomes. These findings highlight the effectiveness of few-shot learning for semantic segmentation and the potential of zero-shot learning in enhancing classification models for knee osteoarthritis diagnosis.

## 1   Introduction

Osteoarthritis (OA) is a chronic joint disease associated with gradual deterioration of articular cartilage [1]. Currently, about 6.7% of the global population is impacted by OA, with knee being the most vulnerable site [2]. Projections indicate a 74.9% increase in knee OA incidence by 2050, leading to a substantial rise in the demand for patient care within this population [2, 3]. The manifestation of knee OA is heterogeneous, but it typically involves chronic knee pain, joint stiffness, reduced range of motion, and loss of locomotive function. General imaging biomarkers of OA comprise articular cartilage damage, osteophyte formation, sclerosis of the subchondral bone, and the development of subchondral cysts [1]. These imaging indicators collectively highlight that the primary site of knee OA is situated in the articulation between the femur and tibia bones.

Knee compartment resembles a hinge-like component that is fundamentally driven by mechanical processes and marked by continuous bone remodeling. The dynamic nature of bone remodeling further emphasizes bone shape as a prospective imaging indicator for OA [4]. Alterations in bone shape or geometry have the potential to influence load distribution within the joint, potentially leading to worsened OA pain or improved OA pain.

In this paper, a simple 2D bone morphological analysis is proposed using manually segmented bones to investigate the morphological characteristics associated with different pain conditions. Furthermore, a deep learning model is proposed for automatic segmentation of knee bones in X-ray images, which can be used as a tool to assist the diagnosis of knee OA. Lastly, the segmentation model is integrated into a classification model for pain condition classification.

Our paper is structured as follows: After the introduction of research background in Section I, Section II provides an overview of related works. Section III describes the database and proposed algorithms. Next, Section IV discusses the experimental findings. Lastly, Section V concludes the study.

## 2    Related Work

Semantic segmentation is a fundamental aspect of bone morphological analysis [5, 6]. It involves the process of isolating the entire bone pixels from raw images. Traditional bone segmentation approaches rely on image processing techniques that recognize low-level features, such as edges, shapes, and pixel intensities [7]. A notable limitation in this domain is the incapacity of a single model to universally accommodate all cases, given that bone structure is subject- specific and not generalized.

Deep learning has leveraged the segmentation procedure through increased computational capabilities, allowing high-level features to be extracted. The integration of a statistical shape model (SSM) into a CNN has been employed to extract the bone surface from imaging data [8, 9]. V-Net encoder- decoder architecture has been commonly used for bone segmentation from MRI images [4]. Bone trajectory analysis becomes feasible with the execution of bone morphological assessments at various time points. Calivá, et al. [4] pioneered a contrastive learning-based deep learning pipeline designed to forecast longitudinal changes in bone shape, specifically focusing on the femur, over a span of 72 months. However, the data acquisition for this study is challenging, as it necessitates multiple MRI images to be taken over time.

Two research gaps are identified based on previous works. Firstly, most proposed algorithms are grounded in supervised learning, assuming there is enough labeled data available for model training. However, these algorithms still struggle to overcome the challenges posed by OA studies with limited labeled data. Consequently, few-shot learning and zero-shot learning emerge as promising deep learning approaches for analyzing datasets with sparse labels. Secondly, despite the higher resolution offered by MRI, it is infrequently used in clinical practice for knee OA diagnosis due to cost constraints. Instead, X-ray imaging remains the most accessible imaging option for confirming OA disease. Considering these two factors, our study will specifically focus on applying few-shot learning and zero-shot learning techniques to X-ray images.

## 3    Database and Methods

### 3.1    Database

The data used for this study is anteroposterior knee X-ray images from Osteoarthritis Initiative (OAI) database [10], which is publicly available at https://nda.nih.gov/oai/. Manual annotations were performed on 606 images to delineate the tibia and femur bones. Pain change over the subsequent 12 months was computed by analyzing the patient self-reported pain scale at baseline and 12 months. Pain increments of 2 were categorized as worsened pain, while pain decreases of 2 were deemed indicative of improved pain.

All images were resized into 640 x 640 pixels. The bone contour was identified from the mask images using marching squares algorithm [11]. Each bone shape was computed using circularity and eccentricity indices. Circularity evaluates how round the bone shape is, and it is defined as:

$$Circularity = \frac{4\pi \times Area}{Perimeter^2} \quad (1)$$

where *Area* is the area of segmented bone and *Perimeter* is the perimeter of segmented bone.

Eccentricity quantifies how much the bone shape deviates from a perfect circle, and it is defined as:

$$Eccentricity = \sqrt{1 - \left(\frac{b}{a}\right)^2} \quad (2)$$

where α represents the length of the semi-major axis and *b* represents the length of the semi-minor axis. The value of eccentricity ranges from 0 to 1, where 0 corresponds to a perfect circle, and 1 corresponds to a line segment.

The data was divided into a training set (536 images), a validation set (45 images), and a test set (25 images).

### 3.2 Model Architectures

In this study, six segmentation models were employed: Res-UNet, DeepLabV3Plus, SegFormer, UniverSeg, bounding box-based segment anything (BBOX-SAM), and central point- based segment anything (CP-SAM).

### 3.3 Supervised Learning

Res-UNet, DeepLabV3Plus, and SegFormer are models that rely on supervised learning. Res-UNet (total params: 2.06M) is an encoder-decoder model, where the encoder consists of three blocks, each employing 3x3 depthwise separable convolutions, followed by batch normalization and residual connections. The decoder comprises four convolutional layers with transposed convolutions, batch normalization, ReLU activation, upsampling layers, and residual connections. Unlike Res-UNet, DeepLabV3Plus [12] (total params: 11.8M) utilizes multi-scale features yielded by the last convolutional block before upsampling. This operation is implemented by atrous spatial pyramid pooling layers. Our DeepLabV3Plus uses ResNet50 as a backbone model. SegFormer [13] (total params: 3.72 M) is a transformer-based segmentation model that utilizes self-attention mechanism. All three models were trained using pixel-wise sparse categorical cross-entropy loss function $L_{pixel-wise-sparse}$ defined as:

$$L_{pixel-wise-sparse}(P, Y) = -\frac{1}{N}\sum_{i=1}^{N} log\ (P_{i,Y_i}) \quad (3)$$

where *P* is probability distribution, *Y* is ground truth label, *N* is the total number of pixels, $Y_i$ is the number of classes, $P_{i,Y_i}$ is the predicted probability of pixel *i* belonging to the true class.

### 3.4 Few-Shot Learning

UniverSeg [14] (total params: 1.18M) is a few-shot learning model that employs a cross-block mechanism to facilitate feature transfer between the support set and the query image. This mechanism adopts an encoder-decoder architecture, where the encoder path consists of multiple levels of CrossBlock and spatial down-sampling operations applied to both the query and support set representations. Conversely, in the expansive path, each level incorporates up-sampling for both representations, effectively doubling their spatial resolutions. The resulting up-sampled representations are then concatenated. Lastly, a single 1x1 convolution is applied to map the ultimate query representation to a prediction. In this study, we initiated bone segmentation on the test set by randomly selecting 31 image-mask pairs from the training set to serve as a support set.

### 3.5 Zero-Shot Learning

Segment Anything (SAM) [15] (total params: 632M) is a transformer-based zero-shot learning model that can be integrated with a customized prompt encoder for segmenting specific objects of interest. The model has been trained on SA- 1B dataset. The notable advantage of this model is its training- free nature. In this study, we developed two prompt encoders: bounding box (BBOX) and central point (CP) (Fig. 1) using AutoML in Roboflow Train 3.0 technology.

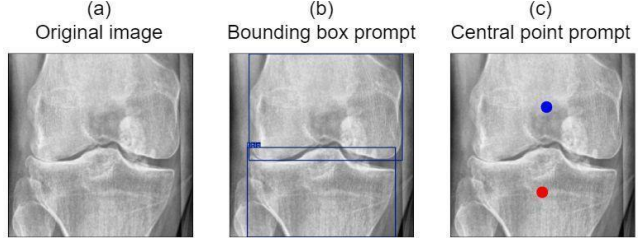

**Fig. 1.** Illustration of (a) original image, (b) bounding box prompt encoder, and (c) central point prompt encoder.

### 3.6 Classifiers and Evaluation Metrics

The classifier for segmentation models is a multilayer perceptron (MLP), while the classifier for categorizing pain conditions is the k-nearest neighbors (KNN) algorithm.

Quantitative evaluation was performed on test set using five metrics, namely accuracy (Acc), precision, recall, Dice score, and Intersection over Union (IoU) derived from true positive (TP), true negative (TN), false positive (FP), and false negative (FN):

$$Acc = 100 \times \frac{TP+TN}{TP+TN+FP+FN} \qquad (4)$$

$$Precision = 100 \times \frac{TP}{TP+FP} \qquad (5)$$

$$Recall = 100 \times \frac{TP}{TP+FN} \qquad (6)$$

$$Dice\ score = 100 \times \frac{2TP}{2TP+FP+FN} \qquad (7)$$

$$IoU = 100 \times \frac{TP}{TP+FP+FN} \qquad (8)$$

## 4 Results

The circularity and eccentricity of femur and tibia bones were computed based on the contour constructed from the masks (Fig. 2). The circularities exhibited variations but generally remained consistent across categories for both bones (Fig. 3(a)). In terms of eccentricity (Fig. 3(b)), both femur and tibia showed a decrease in mean values for the "Worsened pain" category, indicating a shift towards a more circular shape. Femur exhibited a more noticeable decrease, suggesting a pronounced change in bone morphology. In contrast, the tibia showed a slight increase in mean eccentricity for the "Improved pain" category. These findings suggest that bone morphology, particularly in the femur, undergoes notable changes in cases where pain worsens, while improvements in pain might not exhibit pronounced alterations in bone shape.

Based on the segmentation outcomes of femur bones (Table 1), UniverSeg achieved the highest accuracy (99.50%), IoU (99.37%), and Dice score (99.69%), demonstrating superior performance in accurately delineating bone regions. BBOX-SAM and DeepLabV3+ also exhibited strong segmentation outcomes, with accuracies of 97.40% and 98.90%, respectively. However, UniverSeg surpassed them in terms of IoU. In contrast, CP-SAM lagged behind in accuracy (91.12%) and IoU (89.71%), despite achieving the highest precision (99.70%).

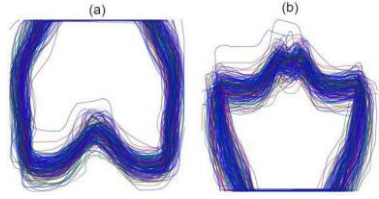

**Fig. 2.** Contours of (a) femur and (b) tibia bones. Blue line represents bone without pain change, red line signifies bone with worsened pain in the next 12 months, and green line indicates bone with improved pain in the next 12 months.

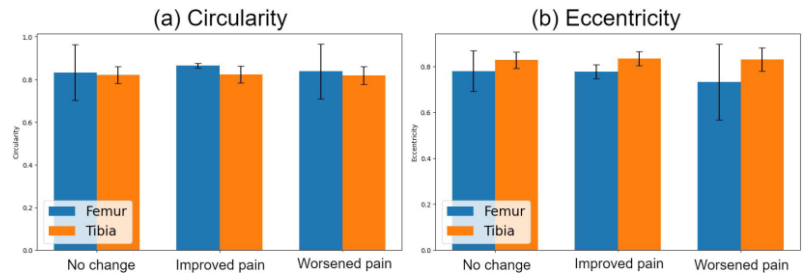

**Fig. 3.** Graphs (a) circularity and (b) eccentricity of femur and tibia bones.

For segmentation of tibia bones (Table 2), UniverSeg emerged as the top performer with the highest accuracy (99.53%), IoU (99.20%), and Dice score (99.60%).

DeepLabV3+ and SegFormer also demonstrated robust segmentation, with accuracies above 98%, IoU scores above 95%, and Dice scores above 97%. CP-SAM, while showing lower accuracy (96.46%), maintained a respectable IoU (87.20%). These findings underline UniverSeg's consistent excellence, especially in capturing the intricate details of bone structures.

Qualitative assessment has been performed (Fig. 4). Res- UNet and SegFormer misidentified the screw on the femur bone as the tibia bone, suggesting that these models are less sensitive to visual artifacts caused by implants. The segmentation quality of BBOX-SAM and CP-SAM was notably influenced by the prompt. In BBOX-SAM, bone segmentation tended to exclude the screw. In CP-SAM, segmentation focused on the point specified in the prompt, and as most central points were in the screw pixel, only the screw part was segmented.

The segmented masks for femur and tibia bones were categorized according to pain conditions using a KNN classifier (Table 3). However, due to the limited dataset size and highly imbalanced class distribution, our classification model exhibited limited effectiveness, achieving an accuracy of only about 52% using ground truth masks. When the classification model was integrated with the segmentation model, only UniverSeg and CP-SAM showed improvements in the classification outcomes by 8% and 14%, respectively.

Given the superior performance of UniverSeg in segmenting bone pixels from X-ray images and its capability to enhance classification model performance, we recommend UniverSeg as the solution for automatic knee bone segmentation. However, the potential of SAM models should not be overlooked, as the prompt encoder can be modified and integrated with domain knowledge to further improve segmentation and classification outcomes.

**Table 1.** Segmentation Outcomes of Femur Bones

| Model | Acc | Precision | Recall | Dice score | IoU |
|---|---|---|---|---|---|
| Res-UNet | 94.24 | 95.69 | 93.47 | 94.57 | 89.70 |
| DeepLabV3+ | 98.90 | 99.03 | 98.76 | 98.90 | 97.82 |
| SegFormer | 98.23 | 99.02 | 98.00 | 98.51 | 97.06 |
| UniverSeg | **99.50** | 99.37 | **100.00** | **99.69** | **99.37** |
| BBOX-SAM | 97.40 | 99.44 | 97.05 | 99.23 | 96.52 |
| CP-SAM | 91.12 | **99.70** | 89.95 | 94.57 | 89.71 |

**Table 2.** Segmentation Outcomes of Tibia Bones

| Model | Acc | Precision | Recall | Dice score | IoU |
|---|---|---|---|---|---|
| Res-Unet | 95.58 | 91.33 | 92.72 | 92.02 | 85.22 |
| DeepLabV3+ | 98.73 | 98.33 | 97.90 | 98.11 | 96.30 |
| SegFormer | 98.73 | 97.48 | 97.91 | 97.70 | 95.50 |
| UniverSeg | **99.53** | **99.20** | **100.00** | **99.60** | **99.20** |
| BBOX-SAM | 97.23 | 96.28 | 95.44 | 95.86 | 92.05 |
| CP-SAM | 96.46 | 92.18 | 94.16 | 93.16 | 87.20 |

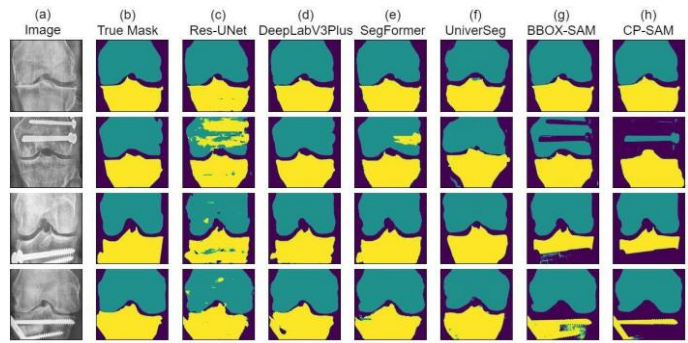

**Fig. 4.** Qualitative assessment of knee bone segmentation. Green masks represent femur bones. Yellow masks represent tibia bones.

**Table 3.** Classification Outcomes of Knee Pain Status

| Model | Acc | F1 |
|---|---|---|
| Baseline – without segmentation | 44 | 45 |
| Ground truth masks | 52 | 55 |
| Res-Unet | 52 | 51 |
| DeepLabV3+ | 52 | 55 |
| SegFormer | 50 | 52 |
| UniverSeg | 60 | 56 |
| BBOX-SAM | 52 | 54 |
| CP-SAM | **66** | **63** |

## 5   Conclusions

In this study we have demonstrated a 2D bone morphological analysis, compared the segmentation capability of six deep learning algorithms, and illustrated how the segmentation model can enhance the classification model. Despite the limited image dataset, the proposed model exhibited effective segmentation results. This model serves as a promising solution in advancing a fully automated diagnostic tool for knee OA. In our future work, we aim to develop an end-to-end algorithm to automatically (i) segment knee bones in MRI sequences, (ii) construct the bone structure into 3D geometry, and (iii) predict the progressive bone changes induced by subject-specific mechanical load.

**Acknowledgment.** This work is supported by the Ministry of Higher Education, Malaysia under Fundamental Research Grant Scheme (FRGS) Grant No. FRGS/1/2022/SKK01/UM/02/1.